\documentclass[aps,prb,twocolumn,superscriptaddress,10pt]{revtex4-2}
\usepackage[utf8]{inputenc}
\usepackage{libertine}
\usepackage[libertine]{newtxmath}
\usepackage{mathtools,graphicx,microtype,bm}
\usepackage[cal=boondoxo,frak=boondox]{mathalfa}
\usepackage[colorlinks=true,allcolors=blue]{hyperref} 

\begin{document}
\title{Interband plasmons due to Mexican hat dispersion in two-dimensional materials with inverted bands}

\author{Vladimir A.\ Sablikov}
\email[E-mail:]{sablikov@gmail.com} 
\affiliation{Kotelnikov Institute of Radio Engineering and Electronics, Fryazino Branch, Russian Academy of Sciences, Fryazino, Moscow District, 141190, Russia}

\begin{abstract}
Two-dimensional topological materials with Mexican-hat dispersion exhibit numerous nontrivial features, such as double-valued isoenergetic contours in $k$-space, strong mixing of electron and hole states, a van Hove singularity in the density of states, and specific quantum geometric properties. We find that, due to these features, Mexican-hat dispersion significantly strengthens interband plasmons, reduces their damping, and gives rise to an additional branch of the plasmon spectrum  when electron-hole symmetry is broken. The plasmon spectra are limited to a finite range of the wave vector, the boundaries of which depend on the strength of the electron-electron interaction. We also reveal a relationship between the features of the plasmon spectrum and singularities of the joint density of states with a finite plasmon wave vector $q$, as well as features of the interband quantum metric, which also depend on $q$.
\end{abstract}

\maketitle

\section{Introduction}\label{S_intro}
Plasmons are collective excitations of the electron liquid~\cite{pines2018theory,giuliani2008quantum} caused by its polarization,which occur in many materials with a wide variety of orbital structures of electron states. Since polarization in modern materials with multi-orbital quantum states is closely related to their quantum-geometric properties~\cite{vanderbilt2018berry}, one might expect that quantities such as quantum metric and connectivity should manifest themselves in the properties of plasmonic excitations. However, it is clear that these are not the only important factors. Band dispersion, the topology of isoenergetic surfaces or contours in $\bm{k}$-space, the valley structure, and the density of states related to the excitations under study also play a decisive role in shaping the properties of plasmons in specific materials. The question of which properties of the basis quantum states generate the specific features of plasmonic excitations in a material with multi-orbital quantum states has attracted much interest and is currently being extensively studied.

Plasmonic excitations of two-dimensional (2D) quantum materials with nontrivial electronic states have been studied in a variety of systems, revealing the specific plasmon properties associated with the specific features electronic states in these materials. Without claiming to be exhaustive, it is sufficient to cite studies of 2D Dirac materials with gapless spectrum~\cite{PhysRevB.87.235418,Ju2011graphene}, Dirac materials with strong spin-orbit interaction~\cite{Stauber_2014}, topological insulators~\cite{PhysRevLett.112.076804,PhysRevB.90.115425,PhysRevB.103.115116,PhysRevB.107.155414}, 2D materials with inverted bands~\cite{PhysRevLett.119.266804}, coupled massless 2D Dirac electron systems~\cite{PhysRevB.85.085443}, twisted bilayer graphene~\cite{ni2015plasmons,hesp2021observation,PhysRevB.106.155422}, atomically thin quasi-2D metals~\cite{da2020universal,do2025slow}, materials described within the framework of the $\alpha$-$T_3$ model~\cite{PhysRevB.105.245414,iurov2021tailoring}, semi-Dirac materials~\cite{PhysRevB.111.045413}. New properties of plasmons are also of significant interest for potential applications~\cite{yan2013damping,rivera2020light,PhysRevLett.111.247401,PhysRevLett.121.086804,wang2021density}.

Plasmonic excitations are classified based on the nature of the electron density fluctuations that form a self-sustaining regime of collective excitations. In accordance with the electronic transitions involved in the formation of fluctuations, plasmons are divided into intraband and interband. Intraband plasmons are generated by fluctuations near the Fermi surface in the allowed band. They have a gapless spectrum. Plasmons of this type have been best studied and offer the greatest potential for practical applications, as they can be quite long-lived in real-world situations~\cite{LIANG2021294}. 

Density fluctuations arising from the electron transitions between different bands generate interband plasmons (IBPs). Their spectrum is characterized by the presence of a gap corresponding to a threshold energy for such transitions~\cite{PhysRevB.75.205418,PhysRevB.80.245435,PhysRevLett.112.076804}. Properties of IBPs are determined not only by the band dispersion, but also, to a significant extent, by the Bloch orbitals that form the basis states in involved bands, as well as their quantum geometric properties. For this reason, it is expected that a variety of new properties of plasmonic excitations due to the peculiarities of electronic states can appear in quantum materials.

Interband plasmons have been studied significantly less than intraband ones. However, they are very attractive because the effects due to quantum geometry and, in general, the multi-orbital structure of electronic states can manifest themselves most strongly in the properties of plasmons of this type. 

Most attention has been paid to IBPs in Dirac electron systems with a gapless spectrum, where IBPs play a significant role at a sufficiently high temperature~\cite{PhysRevB.87.235418,PhysRevB.75.205418,PhysRevB.80.245435}, but become overdamped as $T\to 0$. 

In band gap materials, IBPs can arise even in intrinsic materials when the Fermi level is located inside the band gap. However, the question of the existence of the IBPs and their damping is, generally speaking, not obvious and requires special study for a specific material. Suffice it to say that in simple model systems with a quadratic or linear band spectrum, the IBPs are absent. General criteria for the emergence of IBPs in materials with a more complex band spectrum have not yet been formulated. Interestingly, in the case where the band states can be considered as a mixture of Dirac and Schrödinger states, the IBPs appear, and their damping can be quite weak~\cite{PhysRevLett.112.076804}. Such mixed states are realized within the framework of the BHZ model~\cite{BHZ} in the topological phase with inverted electron and hole bands in the case, when the hybridization of the bands is sufficiently strong.

Theoretical studies of the IBPs using various model approaches have led to the important conclusion that the Landau damping of the IBPs can be strongly suppressed due to strong interband correlations at a sufficiently strong electron-electron (e-e) interaction, when the plasmon energy exceeds the threshold energy of the electron-hole (e-h) continuum~\cite{Stauber_2014,lewandowski2019intrinsically,PhysRevB.103.115116}. The strong enhancement of the interband overlap functions and plasmonic energies can be used to access regimes where Landau damping is entirely quenched, thus enabling applications based on non-dissipative light-matter coupling. 

However, a sufficiently clear understanding of the physical conditions for the existence of IBPs, specific features of their spectra, and damping has not yet been achieved. Some progress in addressing this issue was recently proposed in Ref.~\cite{PhysRevB.106.155422}, which demonstrated that a necessary condition for the existence of IBPs is the presence of a van Hove singularity in the joint density of states (JDOS) and a finite interband Berry curvature for the pair of bands that form this singularity. 

Along with the singularity of JDOS, the features of the overlap function also influence the IBPs. The main effect is that the increase of the overlap function due to band inversion leads to stronger excitation of IBPs~\cite{PhysRevLett.119.266804}.

In this paper, we explore a situation in which a rather complex Mexican-hat-type band dispersion is combined with a nontrivial topology of band states. This model system allows us to determine what new properties the IBPs acquire due to the Mexican hat dispersion, as well as to understand which physical quantities play a key role in the strengthening of the plasmons that occurs in this model, and in the rearrangement of plasmon spectrum.

To be specific, we study 2D topological insulator with Mexican-hat dispersion (MHD) arising due to the inversion of the electron and the hole bands. This fairly universal mechanism of MHD is described within the framework of BHZ model~\cite{BHZ} in the case where the hybridization of the electron and hole bands is not too strong. 

In this case, the electron system possesses a bunch of non-trivial features inherent to it that open up the possibility of the emergence of new properties of collective excitations. (i) The basis quantum states are topological with a specific quantum metric and sufficiently large Berry curvature~\cite{SABLIKOV2025116213}. (ii) In the energy range between the bottom and the top of the MHD, there are two isoenergic contours in $k$-space and therefore an additional channel of electron transitions between them appears~\cite{SABLIKOV2025116213}, due to which asymmetric spin-dependent scattering of electrons arises~\cite{SHCHAMKHALOVA2026418850}. (iii) The effective mass of the quasi-particles changes the sign as a function of energy on the low-wave vector branch of the MHD (specifically, in the conduction band, the effective mass changes the sign from positive to negative)~\cite{SABLIKOV2023115492,SABLIKOV2023129006}, which obviously significantly affects the screening of the potential~\cite{SABLIKOV2025116213,SHCHAMKHALOVA2025417942}. (iv) The presence of a well-known van Hove singularity of the density of states at the MHD bottom, due to which the role of e–e interaction increases significantly. 

We found that the presence of a MHD significantly enhances the polarizability of the electron liquid due to presence of interband transitions between different isoenergetic contours. As a result, the IBPs become stronger, their frequency increases, and damping decreases. The plasmon spectrum acquires unusual features. The plasmon excitations turn out to exist in a finite range of the wave vector $q$, the boundaries of which depend on the strength of the e-e interaction. The lower boundary arises because the interband overlap function vanishes as $q\to 0$ (accordingly, the quantum metric approaches 1). The upper boundary is determined by strong Landau damping at large $q$. Another nontrivial effect arises in the absence of e-h symmetry. In this case, in addition to the main branch of the plasmon spectrum, another branch appears, which has a much lower frequency and exists over a wider range of $q$. The radical rearrangement of the plasmon spectrum that we found with a change in the system parameters and the main of the spectra features are largely determined by two factors: singularities of the JDOS, which depends on the plasmon wave vector $q$, and the interband quantum metric, which also depends significantly on $q$.

The calculations are carried out within the framework of the BHZ model in the random phase approximation (RPA). In Sec.~\ref{S_General}, we present main equations, study general properties of the interband overlap function that defines the interband quantum metric, and the JDOS depending on $q$. The properties of IBPs in the case where the system has e-h symmetry are studied in Sec.~\ref{S_IBP_eh_symmetry}. Effects due to breaking the e-h symmetry and the properties of the additional (low-frequency) branch of the IBPs are presented in Sec.~\ref{S_eh_asymmetry}. In Sec.~\ref{S_conclusion}, we discuss the main results and draw conclusions.

\section{General equations and model details}\label{S_General}
In this section, we describe the model of a 2D material with MHD arising from band inversion, study the general properties of the interband overlap function and the JDOS as functions of the plasmon wave vector, and present a general formalism used in subsequent calculations.

\subsection{The Hamiltonian}\label{ss_Hamiltonian}

As a specific but fairly general model of a MHD we choose the BHZ model~\cite{BHZ}, according to which the MHD is formed due to \textit{s$p^3$} hybridization of the inverted electron and hole bands. For simplicity we restrict ourselves by considering systems with spatial inversion symmetry where spin is a good quantum number. The Hamiltonian for the spin-up component reads
\begin{equation}\label{eq.Hamiltonian}
    H_{\uparrow}= -D \hat{k}^2+
    \begin{pmatrix}
        M-B \hat{k}^2 & A (\hat{k}_x+i\hat{k}_y)\\
        A (\hat{k}_x-i\hat{k}_y) & -M+B \hat{k}^2\\
    \end{pmatrix},
\end{equation}
with $\hbar \bm{k}$ being momentum. The spin-down Hamiltonian is $H_{\downarrow}(\bm{\hat k})= H_{\uparrow}^*(-\bm{\hat k})$. Here $A$, $B$, $D$, and $M$ are material parameters. Further it will be convenient to use dimensionless quantities. The quantities of the energy dimension are presented in units of $|M|$, distance in units of $|M/B|$, and wave vectors in units of $|B/M|$.

There are two important parameter of the model: $a=A/\sqrt{|B M|}$ and $d=D/|M|$. The first describes the hybridization of the electron and hole bands and determines condition under which the MHD exists $|a|<\sqrt{2}$. The parameter $d$ describes the e-h asymmetry. 

The band dispersion is 
\begin{equation}\label{eq.banddispersion}
   \varepsilon_{\lambda}(k)=-d k^2+\lambda\overline{\varepsilon}_k\,, 
\end{equation}
with $\lambda=\pm 1$ being the band index and
\begin{equation}    
\overline{\varepsilon}_k=\sqrt{(1-\nu k^2)^2+a^2k^2}\,,
\end{equation}
where $\nu=MB/|MB|$. The MHD appears in the topological phase, when $\nu=1$ and the hybridization is not too strong $|a|<\sqrt{2}$. To be more specific, we consider the situation with inverted bands for $M<0$ and $B<0$.

\begin{figure}
    \centerline{\includegraphics[width=1.\linewidth]{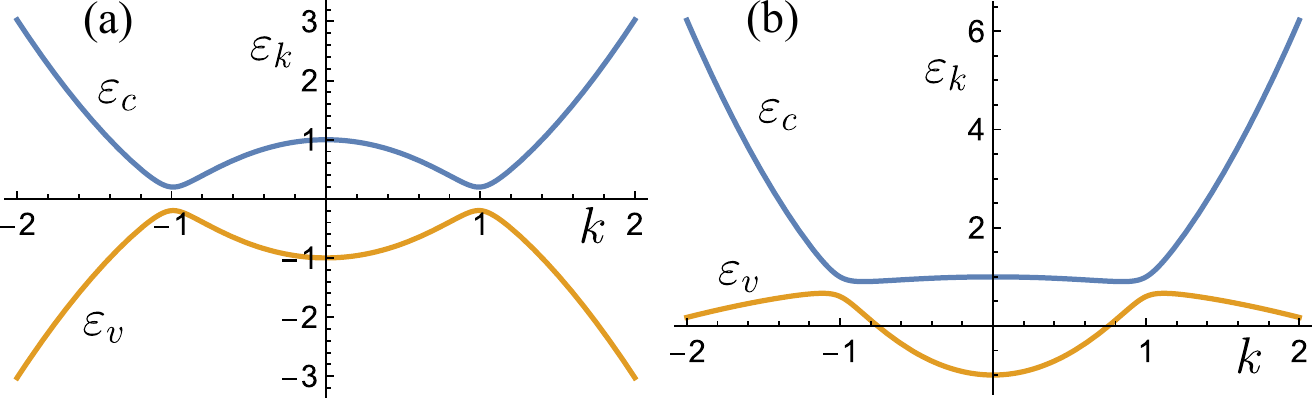}}
    \caption{MHD for the hybridization parameter $a=0.2$ in the cases (a) of e-h symmetric model, $d=0$, and (b) where the e-h symmetry is broken, $d=-0.8$.}
    \label{fig1}
\end{figure}

The MHD is shown in Fig.~\ref{fig1} for two cases: with e-h symmetry and without it. In the first case, the bottom of the MHD $\varepsilon_0=|a|\sqrt{1-a^2/4}$ is reached at $k=k_0=\sqrt{1-a^2/2}$ in both conduction and valence bands. When e-h symmetry is violated, the dispersion bottoms in the conduction and valence bands shift relative to each other.

The basis eigenstates with wave vector $\bm{k}$ are
\begin{equation}
    |s, \lambda, \bm{k}\rangle =\frac{1}{L}u_{s, \lambda}(\bm{k})e^{i\bm{k r}}\,.
\end{equation}
Here $\lambda$ is the band index and $u_{s, \lambda}(\bm{k})$ is a spinor whose rank is determined by the spin and atomic orbitals that form the eigenstates, $L$ is a normalization length.

The spinor $u_{s, \lambda}(\bm{k})$ for spin-up states has the form
\begin{equation}\label{eq.u-spinor}
    u_{\uparrow, \lambda}( \bm{k}) = \frac{1}{\sqrt{2\overline{\varepsilon}_k}}
    \begin{pmatrix}
        \sqrt{\overline{\varepsilon}_k-\lambda (1-k^2)} \\ \dfrac{\lambda ak}{\sqrt{\overline{\varepsilon}_k-\lambda (1-k^2)}} e^{-i\phi}  
    \end{pmatrix}\,,
\end{equation}
where $\phi$ is the polar angle of the vector $\bm{k}$.

\subsection{Polarization function}\label{ss_Polarization_func}
In the study of plasmons, a central role is played by the polarization function, which describes the density-density response of non-interacting electrons~\cite{giuliani2008quantum}. As a function of the wave vector $\bm{q}$ and the frequency $\omega$, the Lindhard polarization function $\Pi^{(0)}(\bm{q},\omega)$ reads~\cite{SABLIKOV2025116213}
\begin{equation}\label{eq.Lidhard_func}
    \Pi^{(0)}\!(\bm{q},\omega)=\sum_{\lambda,\lambda'}\int\! \frac{d^2\bm{k}}{2\pi^2} \frac{n(\varepsilon_{\lambda, \bm{k}})-n(\varepsilon_{\lambda', \bm{k}+\bm{q}})}{\hbar \omega +\varepsilon_{\lambda,\bm{k}}-\varepsilon_{\lambda',\bm{k}+\bm{q}}+i\eta} \mathcal{F}_{\lambda,\lambda'}(\bm{k},\bm{k}+\bm{q})\,,
\end{equation}
where the spin degeneracy is taken into account, and $n(\varepsilon_{\lambda, \bm{k}})$ denotes the occupation factor. 

\subsubsection{Overlapping function}
The form factor $\mathcal{F}_{\lambda,\lambda'}(\bm{k},\bm{k}+\bm{q})$ describes the overlap between the cell periodic parts of the Bloch eigenstates with different quantum numbers:
\begin{equation}\label{eq.overlap_func}
    \mathcal{F}_{\lambda,\lambda'}(\bm{k},\bm{k}+\bm{q})=|u^+_{\lambda}(\bm{k})\, u_{\lambda'}(\bm{k}+\bm{q})|^2\,.
\end{equation}
This function is directly related to the quantum metric of eigenstates~\cite{resta2011insulating}, defined as $D_{\lambda,\lambda'}(\bm{k},\bm{k'})^2=1-|u^+_{\lambda}(\bm{k}) u_{\lambda'}(\bm{k'})|^2$. In the case of the IBPs considered here, the indices $\lambda$ and $\lambda'$ take different values $\lambda \neq \lambda'=\pm 1$. Our study shows that the functional dependence of the overlap function on the electron momentum and the plasmon wave vector helps us understand the origin of characteristic features of the IBP spectrum.

First of all, we note that the $\mathcal{F}_{\lambda,\lambda'}(\bm{k},\bm{k}+\bm{q})$ function vanishes at $q=0$ due to the orthogonality of the periodic parts of the Bloch function cells with the same $k$ in different bands. This, in particular, shows that the interband polarization function Eq.~(\ref{eq.Lidhard_func}) vanishes at $q\to 0$, in contrast to the intraband polarization function.

Direct calculation of the interband overlap function gives
\begin{multline}\label{eq.interband_overlap}
    \mathcal{F}_{\lambda,-\lambda}(\bm{k}, \bm{k}+\bm{q})\equiv \mathcal{F}_2(\bm{k}, \bm{k+q}) \\= \frac{1}{2}-\frac{(1-k^2)(1-|\bm{k+q}|^2)+a^2k(k+q \cos(\phi-\theta))}{2 \sqrt{\overline{\varepsilon}_k \overline{\varepsilon}_{|\bm{k+q}|}}}\,,
\end{multline}
where $\theta$ is the polar angle of the vector $\bm{q}$. For further considerations, it is important to understand how the overlap function varies with $q$. Equation~(\ref{eq.interband_overlap}) clearly shows that $\mathcal{F}_{\lambda,-\lambda}$ is independent of the parameter $d$ and, therefore, does not change when e-h symmetry is broken.  For further considerations, it is important to understand how the overlap function varies with $q$.

For $q\ll 1$, the function $\mathcal{F}_2(\bm{k}, \bm{k}+\bm{q})$ is expanded in powers of $q$
\begin{equation}\label{eq.overlap_q=>0}
   \mathcal{F}_2(\bm{k}, \bm{k}+\bm{q})\approx \frac{a^2}{8}\frac{2 + a^2 k^2 + 2 k^4 + (4 - a^2) k^2 \cos(2 (\phi-\theta))}{[(1 - k^2)^2 + a^2 k^2]^2}q^2 + \cdots \,.
\end{equation}
Thus, the density of the overlap function $\mathcal{F}_2$ is concentrated near a circle of radius $k=k_0$ in $k$-space, where $\mathcal{F}_2$ has a sharp maximum. Depending on the angle $\phi-\theta$, the function $\mathcal{F}_2$ reaches a maximum on the axis parallel to $\bm{q}$ and a minimum in the perpendicular direction. For $q\ll 1$, the function $\mathcal{F}_2$ is proportional to $q^2$.

\begin{figure}
    \centerline{\includegraphics[width=1.\linewidth]{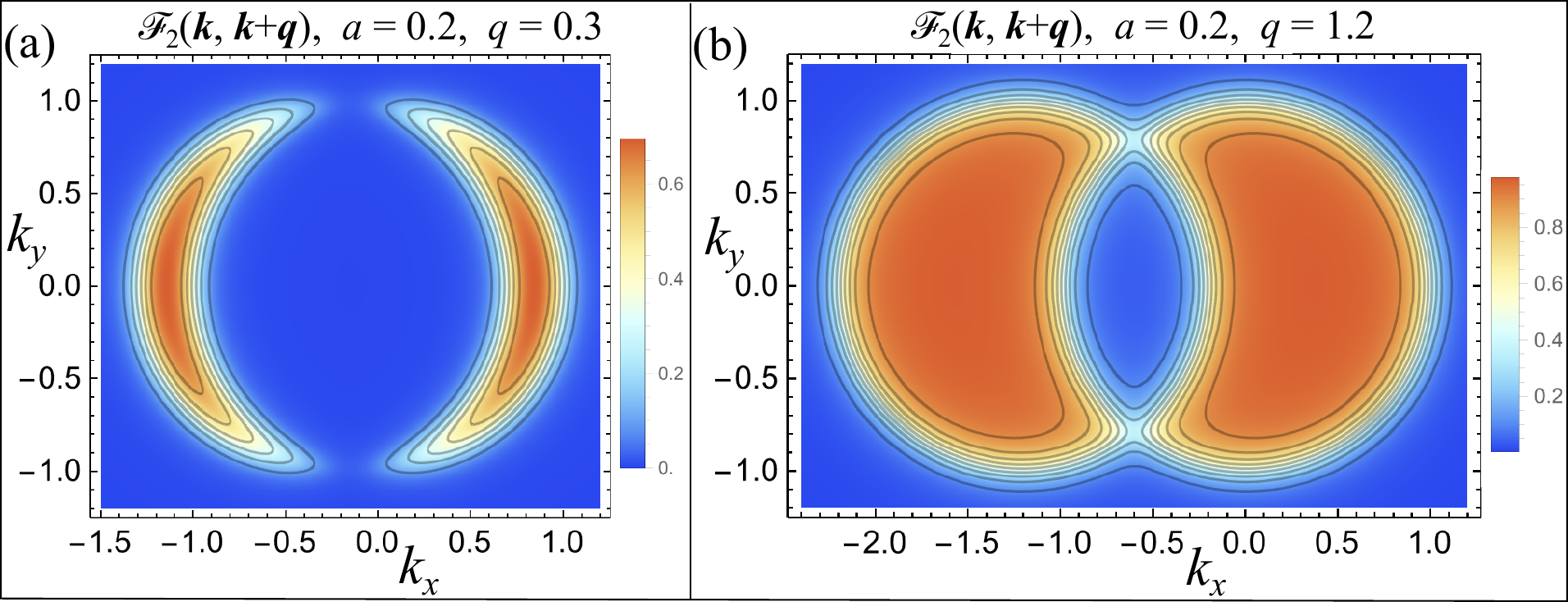}}
    \caption{Distribution of the interband overlap function $\mathcal{F}_2(\bm{k}, \bm{k}+\bm{q})$ in the $\bm{k}$ plane for two values of $q$: (a) $q=0.3$; (b) $q=1.2$. Calculations were carried out for $a=0.2$.}
    \label{fig2}
\end{figure}

As the vector $q$ increases over a wide range, the distribution of the function $\mathcal{F}_2$ on the $\bm{k}$-plane changes, as shown in Fig.~\ref{fig2}. The regions where $\mathcal{F}_2$ reaches a maximum expand, with $\mathcal{F}_2$ increasing almost to unity. In the region between the maxima, $\mathcal{F}_2$ drops sharply to almost zero. As $q$ increases, the overall picture shifts in $k$-space by an amount of the order of $q/2$.

\subsubsection{Joint density of states}\label{s_JDOS}
Another factor that largely determines the polarization function $\Pi^{(0)}\!(\bm{q},\omega)$ is the JDOS, which reflects the fact that electronic excitations with a finite wave vector $\bm{q}$ participate in the polarization process. In the literature, the joint density of states for direct transitions is usually considered when $q=0$~\cite{CABRERA2016103}; this same JDOS is also used in the analysis of plasmon spectra~\cite{PhysRevB.106.155422}. In this paper, we find that it is the JDOS with the finite momentum of electronic excitations that plays an important role in the formation of plasmon spectra, especially when the density of states has singularities. The point is that the structure of singularities, in the case under study, van Hove singularities, can be radically transformed due to a change in the wave vector. In what follows we will use the abbreviation JDOSq for the JDOS with the wave vector $\bm{q}$ of electron excitations.

The JDOSq is determined by the interband excitation energy, 
\begin{equation}\label{eq.interbandenergy}
  w_{\bm{k}+\bm{q},\bm{k}}=\varepsilon_c(\bm{k}+\bm{q})-\varepsilon_v(\bm{k})\,. 
\end{equation} 
Therefore, JDOSq is uniquely determined by the wave vector $\bm{k}$ for a given $\bm{q}$, but in the case of mexican-hat shaped dispersion cannot be represented as a function of energy alone, since $k$ is not a single-valued function of energy, generally speaking, neither in the $c$- nor in the $v$-band.

Nevertheless, a qualitative understanding of the JDOSq singularities can be obtained as follows. Formally, the JDOS appears when calculating the integrals of the form of the polarization function Eq.~(\ref{eq.Lidhard_func}). To do this, we move from integration over $d^2 k$ to integration over the excitation energy $w= w_{\bm{k}+\bm{q},\bm{k}}$ and the angle $\phi'=\phi-\theta$, \begin{equation}\label{eq.JDOS}
    \int\!\!d^2k \mathfrak{F}(\bm{k},\bm{k}+\bm{q}) =\int \!\!dw \underbracket{\sum_{i}\!\int\limits_{-\pi}^{\pi} \frac{d\phi' k}{\Bigl|\frac{\partial w_{\bm{k}+\bm{q},\bm{k}}}{\partial k}\Bigr|}} \mathfrak{F}(\bm{k},\bm{k}+\bm{q})\Biggr|_{k=k_i(w,\phi';q)}\,.
\end{equation}
Here $\mathfrak{F}(\bm{k},\bm{k}+\bm{q})$ denotes an integrant, $k_i(w, \phi; q)$ is a solution of the equation $w=w_{\bm{k}+\bm{q},\bm{k}}$ with respect to $\bm{k}$. The index $i$ numbers these solutions each of which forms a contour in the $\bm{k}$-space. The underlined part represents the JDOSq, which in this case is a kernel of the integral operator.

Singularities of the JDOSq appear when $|\nabla_{\bm{k}} w_{\bm{k}+\bm{q},\bm{k}}|=0$. Thus the singular points can be found from the shape of the surface $w=w_{\bm{k}+\bm{q},\bm{k}}$ in the $\bm{k}$ space for a given $q$, more precisely, saddle points and extrema. At the saddle points the JDOSq diverges logarithmically, and at the extrema the JDOSq undergoes a jump, as in the theory of van Hove singularities~\cite{PhysRev.89.1189}. 

\begin{widetext}
  
\begin{figure}
    \centerline{\includegraphics[width=0.9\linewidth]{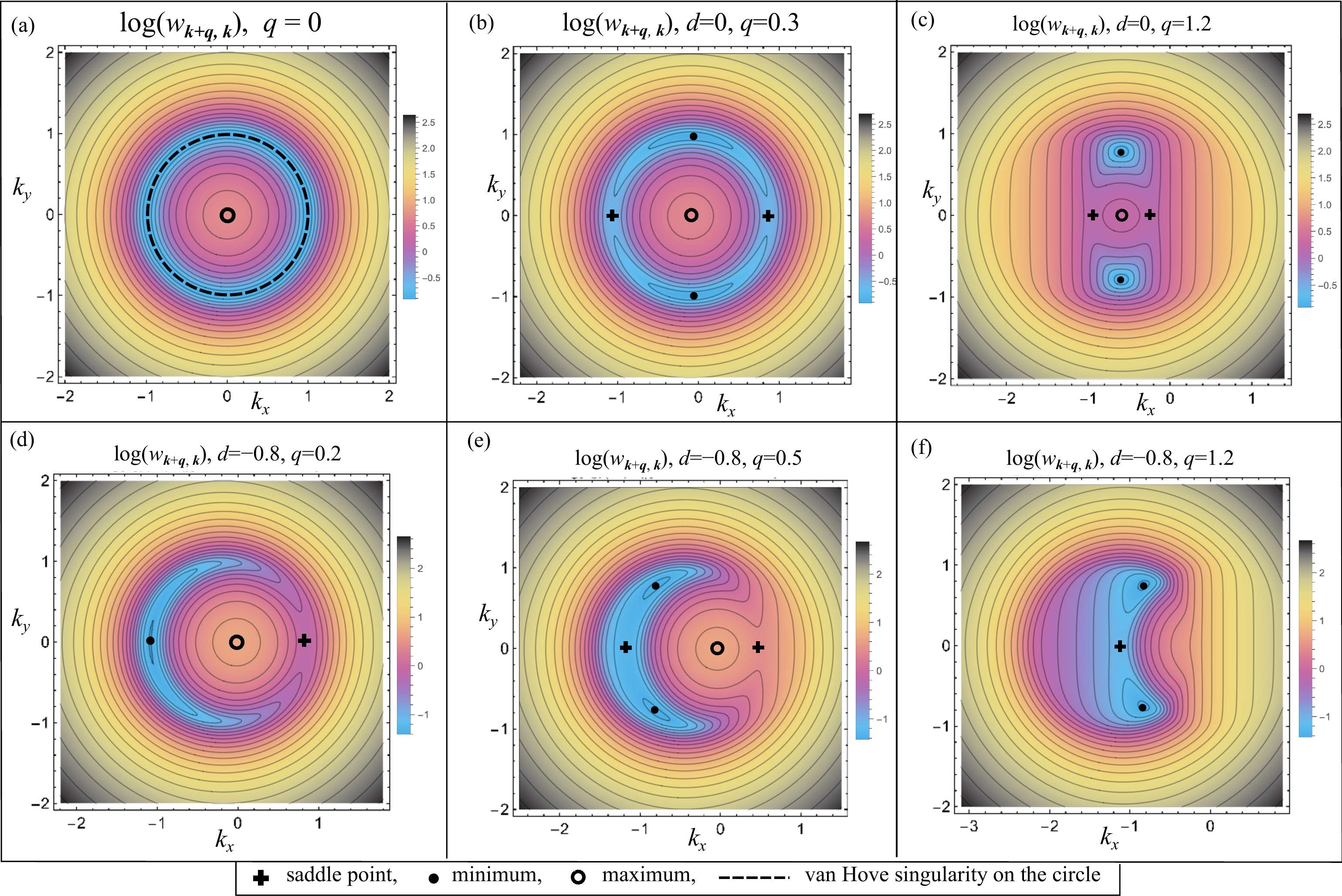}}
    \caption{Map of the surface of the interband energy $w_{\bm{k}+\bm{q},\bm{k}}$ and the evolution of JDOSq singularities with changing $q$ and the e-h asymmetry parameter $d$. (a) The case $q = 0$, when the interband energy is independent of $d$. The dashed line denotes the van Hove singularity on the circle. (b) The case $d=0$ and $q=0.3$. Increasing $q$ leads to a transformation of the van Hove singularity on the circle into two saddle points and two minima. (c) The case $d=0$ and $q=1.2$. The saddle points and minima shift to the center as $q$ increases. (d) The case of e-h asymmetry, $d$=-0.8, for $q=0.2$. Increasing $q$ transforms the van Hove singularity on the circle into single saddle point (hereinafter referred to as high-energy saddle point) and single minimum. (e) The case $d$=-0.8 and $q=0.5$. The minimum of $w_{\bm{k}+\bm{q},\bm{k}}$ splits into two minima separated by a saddle point (low-energy saddle point). (f) The case $d$=-0.8 and $q=1.2$. With increasing $q$ the singularities converge toward the center and the central maximum disappears. Isoline topologies near the singularities indicate their type and location. The calculations were performed for $a=0.2$.}
    \label{fig3}
\end{figure}

\end{widetext}

The evolution of the surface map of the function $w=w_{\bm{k}+\bm{q},\bm{k}}$ with changing $q$ is shown in Fig.~\ref{fig3}. Here, the isolines clearly indicate the presence of extrema and saddle points, which are rearranged when the wave vector $q$ and the e-h asymmetry parameter $d$ change. It is convenient to consider two situations separately: the case of e-h symmetry, $d=0$, shown in panels (a)-(c), and the case when the e-h symmetry is strongly broken, panels (d)-(f). Note that Fig.~\ref{fig3} uses a logarithmic scale for the function $w_{\bm{k}+\bm{q},\bm{k}}$, which improves visual perception.  

Let us first consider the case $q=0$, when the JDOSq does not depend on the e-h asymmetry parameter. In this case, according to Eqs.~(\ref{eq.interbandenergy}) and (\ref{eq.banddispersion}), the interband energy simplifies to $w_{\bm{k}+\bm{q},\bm{k}}=2\overline{\varepsilon}_k$ and, therefore, coincides with the band dispersion in the symmetric model up to a factor of 2. Thus, the interband energy has a well-known van Hove singularity on a circle of radius $k_0$ with logarithmic divergence of the JDOSq and a maximum in the center at the level $w_{0,0}=2$. An example of the interband energy map in this case is shown in Fig.~\ref{fig3}(a). 

At a finite value of $q$, the singularities undergo a significant transformation, which occurs differently in the symmetric and asymmetric cases.

In the presence of e-h symmetry, the singularities transform as follows (see Figs.~\ref{fig3}(a - c)). As $q$ increases, the van Hove singularity on the circle transforms into two saddle points and two local minima. The singularity points are located on the axes symmetrically relative to the center, as shown in Figs.~\ref{fig3}(b,c), where the horizontal axis is parallel to the vector $-\bm{q}$. A further increase of $q$ leads to a shift of the saddle points toward the center. The minima also shift to the center but more slowly, Fig.~\ref{fig3}(c). Overall, the picture maintains symmetry with respect to the horizontal and vertical axises.

If the e-h symmetry is broken, an increase of the wave vector $q$ leads to an asymmetric transformation of the singularities with respect to the axis perpendicular to $\bm{q}$, see Figs.~\ref{fig3}(d - f). As $q$ increases, first of all, the van Hove singularity is transformed into single saddle point and single minimum, located on the horizontal axis to the right and left of the center. A further increase of $q$ leads to another transformation. The minimum splits into two minima separated by a saddle point, Fig.~\ref{fig3}(e). Thus, two minima appear, symmetrically located above and below the horizontal axis, as well as a saddle point on the horizontal axis. The further evolution of singularities with increasing $q$ consists of the convergence of saddle points, which results in the disappearance of the central maximum.

Of great interest is the energy value at singularity points and its dependence on $q$. In the case of e-h symmetry, there are two singularities: a saddle point and a minimum. However, only the saddle point is of interest, since the contribution of the minimum to the susceptibility integral is negligible. This is because the minimum point falls in a region where the overlap function is small. The interband energy $w_s (q)$ at the singularity point increases monotonically with the vector $q$, as can be seen in Figs.~\ref{fig3}\,(b, c).

\begin{figure}
    \centerline{\includegraphics[width=0.8\linewidth]{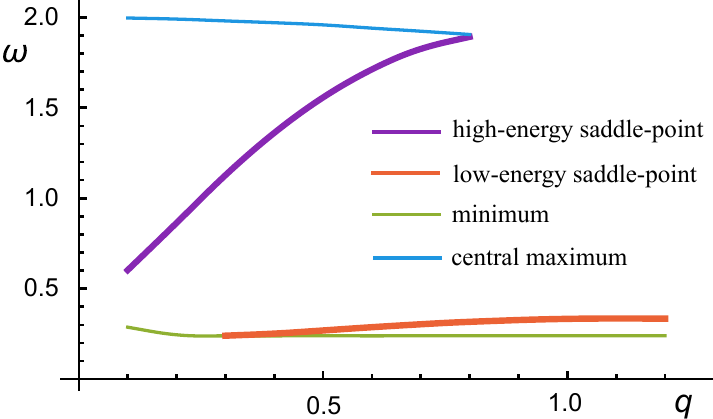}}
    \caption{Energy spectrum of the JDOSq singularities as a function of the wave vector $q$ in the case of broken e-h symmetry with $d=-0.8$ and $a=0.2$. Bold lines are branches of the low- and high-energy saddle-point singularities.}
    \label{fig4}
\end{figure}

The situation changes radically when the e-h symmetry is broken. This is especially important in the context of further studies of plasmon spectra. The spectra of the singularities are shown in Fig.~\ref{fig4} for the same parameters as in Figs.~\ref{fig3}~(d - f). There are four singularities:\\ 1) the minimum of $w_{\bm{k}+\bm{q},\bm{k}}$ (marked by solid dot in Figs.~\ref{fig3});\\ 2) the low-energy saddle point (left cross in panel (e));\\ 3) and the high-energy saddle point (right cross in panel (e));\\ 4) the central maximum (marked by a ring). \\Thus, the singularity spectrum contains four branches shown in Fig.~\ref{fig4}. 

However, not all singularities are equally important for two reasons. First, saddle points contribute more to the integral than extrema. Second, the contribution of a singularity to the susceptibility integral is determined not only by the behavior of the JDOSq, but also by the magnitude of the overlap function in the singularity region, which, as shown above, has maxima and deep minima. Thus, saddle points fall in the regions where the overlap function $\mathcal{F}_2(\bm{k}, \bm{k}+\bm{q})$ is close to its maximum. In contrast, the minima of $w_{\bm{k}+\bm{q},\bm{k}}$, shown in panel (e), are located in the region where $\mathcal{F}_2(\bm{k}, \bm{k}+\bm{q})$ drops to minimum values. And only in a small interval of $q$ in panel (d), the maximum of $w_{\bm{k}+\bm{q},\bm{k}}$ falls in the region where $\mathcal{F}_2(\bm{k}, \bm{k}+\bm{q})$ close to its maximum. As for the central maximum, our calculations show that this singularity, lying in the high-energy region does not play a significant role in the plasmon spectrum at all. Thus, for a qualitative interpretation of plasmon spectra, it can be assumed that the main role is played by two saddle-point singularities: low-energy and high-energy.   

\subsection{Dielectric function}\label{ss_dielectric func}
The dielectric function is calculated within the RPA,~\cite{giuliani2008quantum} 
\begin{equation}\label{eq.dielectric_func}
    \epsilon_{\mathrm{RPA}}(\bm{q}, \omega)= 1-v_{ee}(q)\,\Pi^{(0)}(\bm{q},\omega)\,,
\end{equation}
where $v_{ee}(q)$ is Fourier component of the e-e interaction potential. To be specific, we will henceforth consider the interaction to be Coulombic. In this case, in dimensionless units 
\begin{equation}
    v_{ee}(q)= \frac{Q}{q}\,,\quad \mathrm{with}\quad Q=\frac{2\pi e^2}{\overline{\epsilon}\sqrt{|MB|}}\,,
\end{equation}
where $\overline{\epsilon}$ is effective permittivity of the medium. Thus, the e-e interaction effects are described by the parameter $Q$, which significantly determines the conditions for plasmon excitation, their velocity and decay.

At a qualitative level, collective modes can be visualized as sharp peaks of the energy-loss function in the $q$-$\omega$ plane 
\begin{equation}
    \mathcal{L}(q, \omega)=-\mathrm{Im}\left[\frac{1}{\epsilon_{RPA}(q, \omega)}\right]\,.
\end{equation}
It is important to understand how far this peak is located relative to the boundary of the e-h continuum, where collective modes decay into single pairs due to Landau damping. The e-h continuum occupies the region of the $q$-$\omega$ plane where $\mathrm{Im}\,\Pi^{(0)}(\bm{q},\omega)\ne 0$~\cite{giuliani2008quantum}. 

Collective excitations are studied in more detail by analyzing the zeros of the dielectric function $\epsilon_{RPA}(q, \omega)=0$, which are found by expanding $\epsilon_{RPA}(q, \omega)$ near the zeros of the $\mathrm{Re}\,\epsilon_{RPA}(q, \omega)$, assuming that the damping is sufficiently weak.

Using the approaches described above, in the following sections we will study the dynamic permittivity of interacting electrons and collective excitations in the BHZ model. This will allow us to reveal the effect produced by MHD and determine what new effects arise as a result of the breaking of e-h symmetry.

\section{Interband plasmons with e-h symmetry}\label{S_IBP_eh_symmetry}
We begin with a model that has e-h symmetry. Direct calculation of the dielectric function using Eqs.~(\ref{eq.Lidhard_func}) and (\ref{eq.dielectric_func}) yields the following results.

\subsection{IBP spectrum}
\begin{figure}
    \centerline{\includegraphics[width=1.\linewidth]{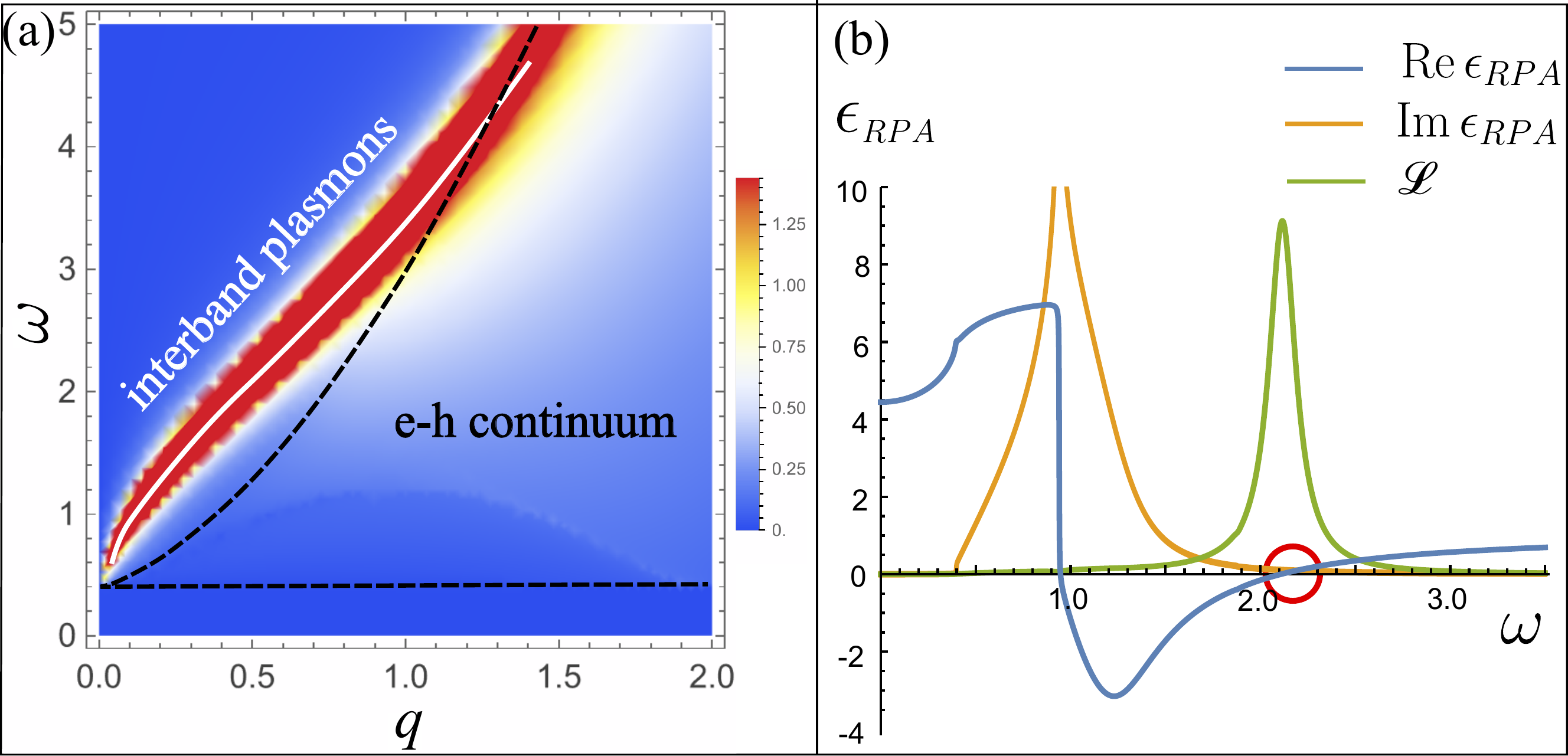}}
    \caption{(a) The distribution of the loss function density in the $q$-$\omega$ plane in the case of the e-h symmetry. The white line represents the IBP dispersion. The black dashed line is the boundary of the e-h continuum. (b) Line plot of the real (blue solid line) and imaginary (dark yellow line) parts of the dynamical dielectric function as well as the loss function $\mathcal{L}(\omega)$ (green line) for $q =0.5$. The zero of $\epsilon_{RPA}(\omega)$ (marked with a red circle) represents the plasmon node. The parameters are $a=0.2$, $d=0$, $Q=10$.} 
    \label{fig5}
\end{figure}

The energy loss function $\mathcal{L}$ has a sharp peak in the $q$-$\omega$ plane, indicating plasmon excitation. A typical example is shown in Fig.~\ref{fig5}(a) for a relatively small hybridization parameter, $a=0.2$, when the ``Mexican hat'' shape is sufficiently deep. The trajectory of the peak of the $\mathcal{L}(q, \omega)$ function shows the plasmon dispersion. For plasmon excitation to be efficient, it is important that the dispersion line is located outside the region of the e-h continuum, where collective excitations decay because of Landau damping. The continuum boundaries are shown by dashed lines.

\begin{figure}%
    \centerline{\includegraphics[width=1.\linewidth]{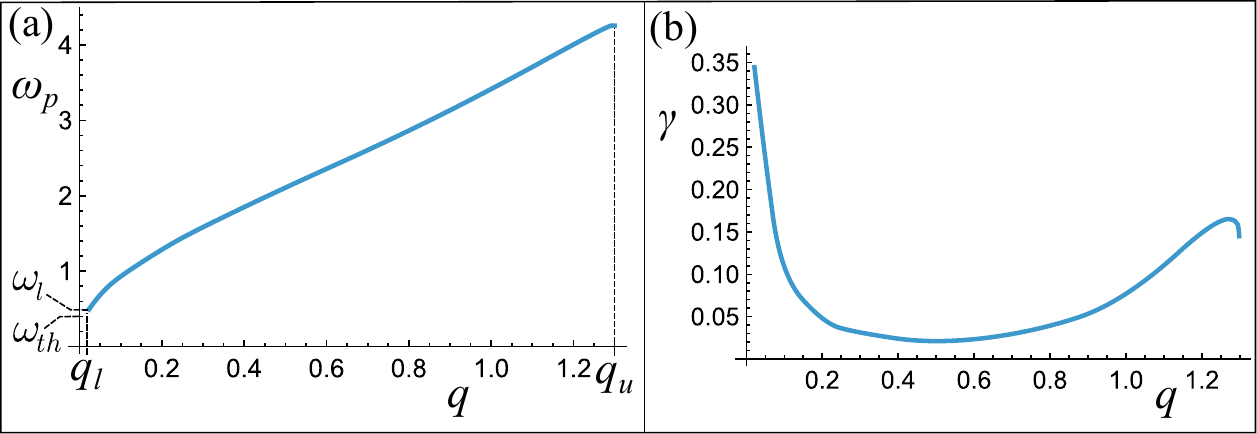}}
    \caption{(a) Dispersion $\omega_p(q)$ of the IBPs and (b) the decay constant $\gamma(q)$ in e-h symmetric case. $q_l$ and $q_u$ are the lower and upper boundaries of the wave vector range where the IBPs are excited. $\omega_l$ is lower frequency and $\omega_{th}$ is the threshold energy of interband transitions. The parameters are $a=0.2$, $d=0$, $Q=10$.}
    \label{fig6}
\end{figure}

A significant portion of the $\mathcal{L}$-peak trajectory is seen to pass outside the e-h continuum region and crosses its boundary at a sufficiently large $q$. In this segment, the plasmons are well defined. In more detail, the plasmon spectrum $\omega_p(q)$ and the damping $\gamma(q)$ are calculated by expanding $\mathrm{Re}\,\epsilon_{RPA}$ and $\mathrm{Im}\,\epsilon_{RPA}$ as functions of $\omega$ near the zeros of $\epsilon_{RPA}(q,\omega)$ on the complex plane $\omega=\omega'+i \omega''$, assuming that the damping is small. The dependencies of the real and imaginary parts of the dielectric function on frequency are shown in Fig.~\ref{fig5}(b). The results of these calculations of $\omega_p(q)$ and $\gamma(q)$ are shown in Fig.~\ref{fig6}(a, b) for the same set of material parameters as in Fig.~\ref{fig5}. 

The plasmon spectrum, Fig.~\ref{fig6}(a), has a form typical for interband excitations. Of considerable interest is the quality factor of plasmonic excitations, which in the case under consideration turns out to be quite large, on the order of $10^2$ as in the example shown in Fig.~\ref{fig6}(b).

But of most interest is the fact that the spectrum of IBPs is confined within a finite range of $q$ that has both lower and upper boundaries, $q_l$ and $q_u$. The lower boundary of the frequency $\omega_l$ is somewhat higher than the threshold energy of direct interband transitions $\omega_l>\omega_{th}=w(k_0,k_0)$. The upper boundary $q_u$ is obviously due to the approaching of the plasmon spectrum to the e-h continuum and subsequent decay of plasmons. The lower boundary $q_l$ has another origin, which can be revealed by considering the susceptibility in the limit $q\to 0$. However, due to the presence of MHD the system has a singularity point $(q=0, \omega=\omega_{th})$, so the expansion in $q$ is strongly complicated. 

We have found that the lower boundary arises as a consequence of the combined action of two specific factors. First, the band overlap function $\mathcal{F}_{2}(\bm{k},\bm{k}+\bm{q})$ tends to zero in the limit $q\to 0$ as $\mathcal{F}_{2}(\bm{k},\bm{k}+\bm{q})\sim q^2$, see Eq.~(\ref{eq.overlap_q=>0}). This is a fundamental property of the interband overlap function which shows that the $\Pi^{(0)}(q,\omega)$ tends to zero if the pole part of the integrand in Eq.~(\ref{eq.Lidhard_func}) does not strongly diverges as $q\to 0$. The second factor is the presence of van Hove singularities on the interband energy surface $w=w_{\bm{k}+\bm{q},\bm{k}}$, which are strongly transformed with a change in $q$ for $q<<1$, as shown in Fig.~\ref{fig3}(a, b, d). It is this factor that determines the dependence of the pole contribution to $\Pi^{(0)}(q,\omega)$ on $q$. Therefore, band dispersion must be carefully taken into account when calculating $\Pi^{(0)}(q,\omega)$. Unfortunately, such calculations cannot be performed completely analytically even by expanding the $\mathcal{F}_{2}(\bm{k},\bm{k}+\bm{q})$ and $w_{\bm{k}+\bm{q},\bm{k}}$ functions in $q$.

\begin{figure}[h]
    \centerline{\includegraphics[width=0.95\linewidth]{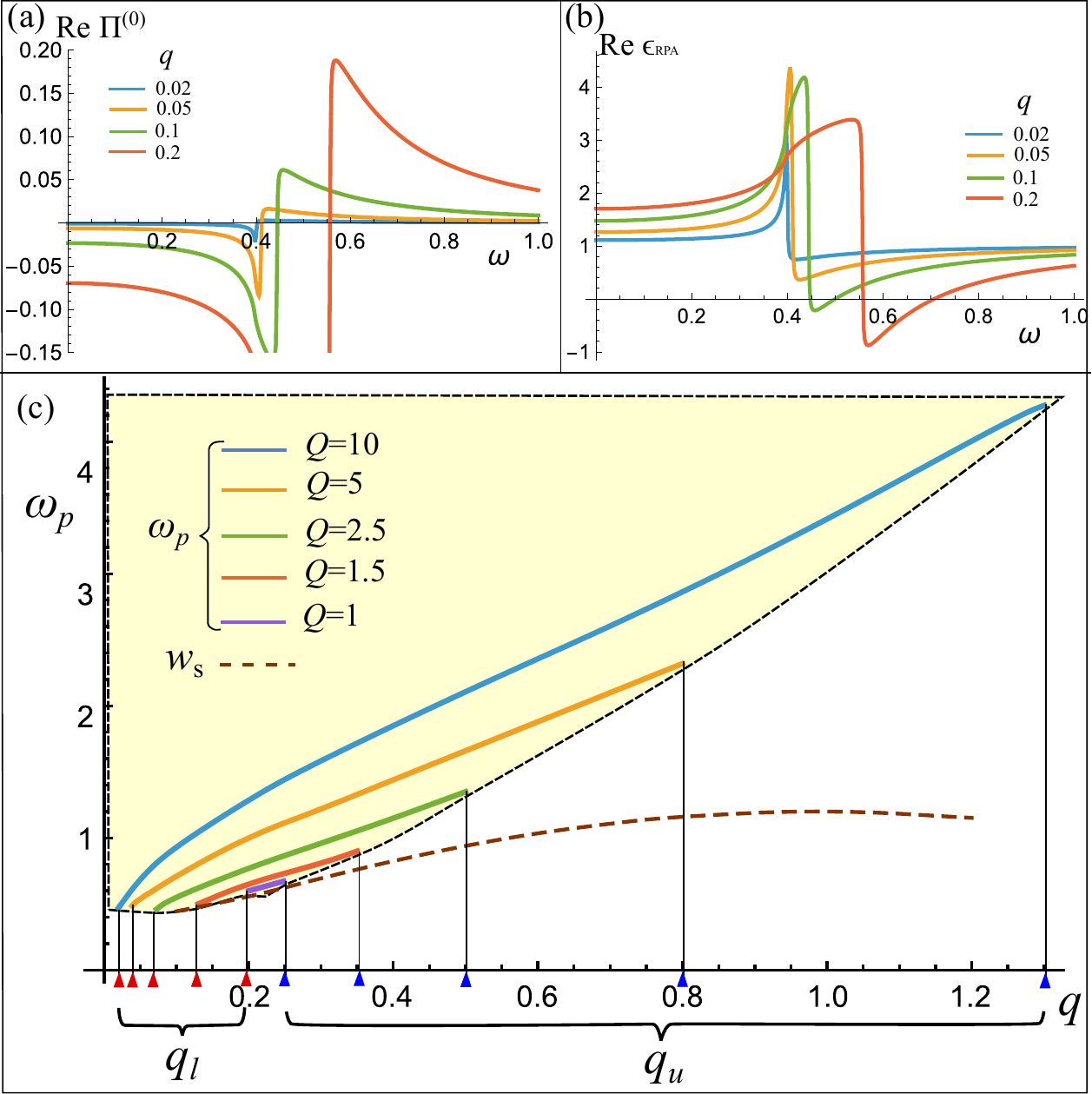}}
    \caption{(a) Screening function $\mathrm{Re}\,\Pi^{(0)}(q, \omega)$ and (b) Real part of the dielectric function $\mathrm{Re}\,\epsilon_{RPA}(q, \omega)$ as functions of $\omega$ for a set wave vector $q\ll 1$. The parameters are $a=0.2$, $d=0$, $Q=2$. (c) Dispersion $\omega_p(q)$ of the IBPs in the case of the e-h symmetry for a set of the interaction parameter $Q$. $q_l$ and $q_u$ are the lower and upper boundaries of the spectrum. Bold dashed line is the energy $w_s(q)$ of the saddle-point singularity. Thin dashed line is lower boundary of the region in the $q$-$\omega$ plane (shaded in yellow) where the IBP spectrum exists for $Q>Q_c$. The parameters are $a=0.2$, $d=0$.}
    \label{fig7}
\end{figure}

Numerical calculations of $\Pi^{(0)}(q,\omega)$ for $q\ll 1$ allow us to identify the reason for the occurrence of the lower boundary $q_l$ of the plasmon spectrum. If the band dispersion is correctly taken into account, the dynamic screening function changes sign with increasing $\omega$, demonstrating the antiscreening required for plasmon generation. However, its maximum positive value decreases as $q\to 0$. According to numerical calculations this dependence is roughly approximated by a power function $q^\nu$ with $\nu \approx 2$. As a result, the plasmonic node of the dielectric function, $\mathrm{Re}\,\epsilon_{RPA}(q,\omega)=0$, disappears with decreasing $q$. Zeros of $\mathrm{Re}\,\epsilon_{RPA}(q,\omega)$ appear only when $q$ exceeds a critical value $q=q_l$. An example of such a behavior is shown in Fig.~\ref{fig7}(a, b). for two values of the interaction constant $Q$. It is worth noting that if the band dispersion is neglected or simplified too roughly, then zeros of $\mathrm{Re}\,\epsilon_{RPA}(q,\omega)$ appear even at $q\ll 1$~\cite{PhysRevB.106.155422}.

The presence of an anti-screening maximum $\mathrm{Re}\,\Pi^{(0)}_{max}$ of the screening function $\mathrm{Re}\,\Pi^{(0)}(q, \omega)$, as in Fig.~\ref{fig7}(a), suggests that plasmons cannot be excited when the e-e interaction strength is not strong enough. This follows from Eq.~(\ref{eq.dielectric_func}). If the interaction parameter $Q$ is lower than a critical value $Q_c\sim q\,\mathrm{Re}\,\Pi^{(0)}_{max}$ the dielectric function may have no zeros for a given $q$. It also follows from this that the boundaries $q_l$ and $q_u$ of the interval in which plasmons exist depend on the strength of the e-e interaction described by the parameter $Q$.

These qualitative estimates prompted us to study in more detail the influence of e–e interactions on the plasmon spectrum. To this end, we calculated the plasmon frequency directly by analyzing the complex dielectric function $\epsilon_{RPA}(q, \omega'+i\omega'')$ near the zeros of $\mathrm{Re}\,\epsilon_{RPA}$ for a wide range of $Q$. The result is presented in Fig.~\ref{fig7}(c) for the hybridization parameter $a=0.2$. For completeness, the energy of the saddle-point singularity of the JDOSq is also presented.

It is evident that the spectrum $\omega_p(q)$ is defined in a finite interval of the wave vector $q_l<q <q_u$. Formally, the boundaries arise because the equation $\mathrm{Re}\,\epsilon_{RPA}(q, \omega)=0$ with a given $q$ has no real roots when $q$ is sufficiently small or sufficiently large. As the parameter $Q$ increases, the lower and upper boundaries $q_l$ and $q_u$ move apart and the frequency increases. Conversely, as $Q$ decreases, the interval $[q_l, q_u]$ collapses to a point at some minimum value of $Q=Q_c$. Interestingly, at this point the frequency is close to the energy of the saddle-point singularity. At $Q<Q_c$ plasmons are absent.

\subsection{Plasmon enhancement due to MHD}
Now we turn to the question of the role of MHD in the excitation of IBPs. The IBPs are known to occur at zero temperature when the band dispersion is more complex than linear or quadratic, and, accordingly, the band states are intermediate between Dirac and Schrödinger states~\cite{PhysRevLett.112.076804}. Within the BHZ model, this situation occurs at $a>\sqrt{2}$. In the Mexican hat regime, at $a<\sqrt{2}$, two dispersion branches are formed, two isoenergetic contours in the $\bm{k}$-space appear with the same energy, and, accordingly, new channels of interband electron transitions appear between two isoenergetic contours in the $v$-band and two contours in the $c$-band in different combinations. Therefore, one can expect that the electronic polarizability will strongly increase and, accordingly, plasmons will be strengthened.

To study the possibility of plasmon excitation, it is important to first analyze the real part of the polarization function, which describes screening. A necessary condition for plasmon excitation is a change in the sign of $\mathrm{Re}\,\Pi^{(0)}(q, \omega)$. In this case, the screening that occurs when $\mathrm{Re}\,\Pi^{(0)}<0$ is replaced by anti-screening when $\mathrm{Re}\,\Pi^{(0)}>0$. The real part of the dielectric function $\mathrm{Re}\,\epsilon_{RPA}=1-(Q/q)\,\mathrm{Re}\,\Pi^{(0)}(q,\omega)$ can vanish only when $\mathrm{Re}\,\Pi^{(0)}>0$.

\begin{figure}
    \centerline{\includegraphics[width=1.0\linewidth]{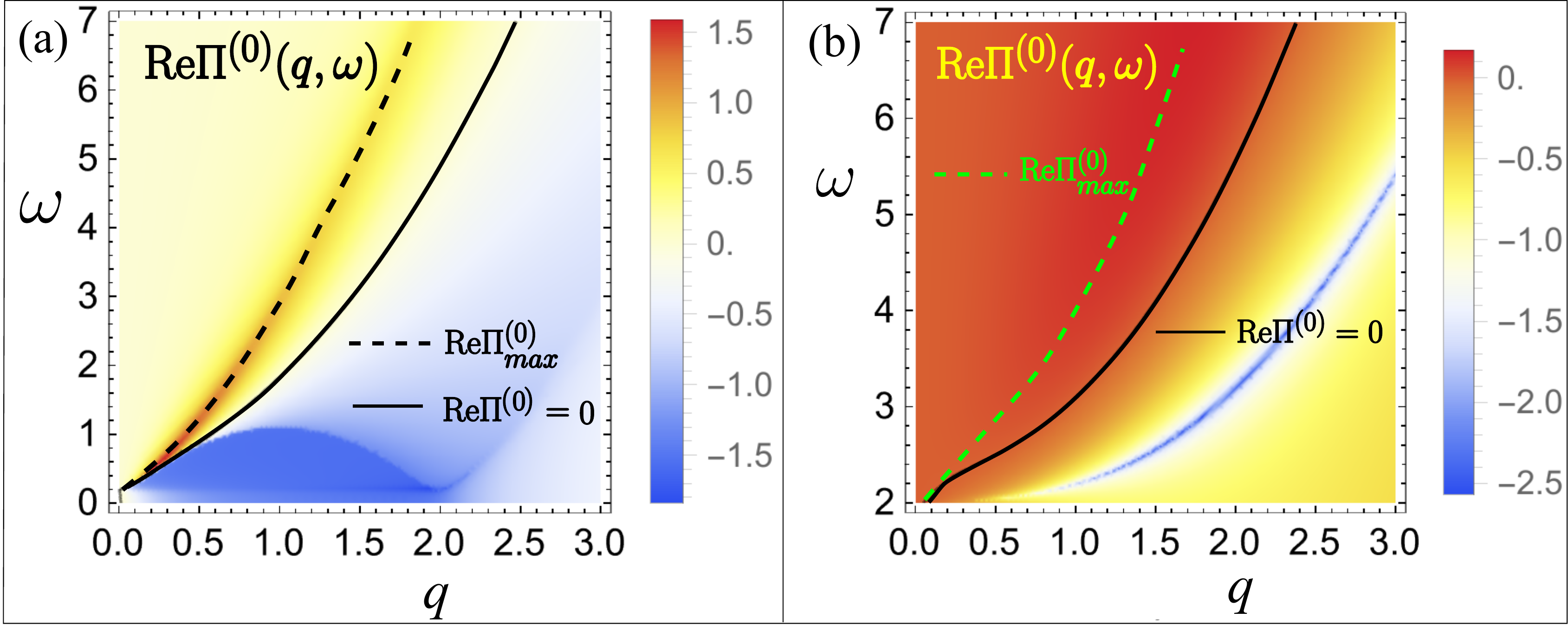}}
    \caption{(a) The distribution of the screening function $\mathrm{Re}\,\Pi^{(0)}$ on the $q$-$\omega$ plane in the case of MHD for $a=0.1$ and $d=0$. (b) The screening function $\mathrm{Re}\,\Pi^{(0)}\!(q,\omega)$ in the absence of MHD for $a=1.6$ and $d=0$. In both cases the solid lines show the trajectory of the zeros of $\mathrm{Re}\,\Pi^{(0)}\!(q,\omega)$ on the $q$-$\omega$ plane. In the MHD case the maximum value of $\mathrm{Re}\,\Pi^{(0)}\!(q,\omega)=\mathrm{Re}\Pi^{(0)}_{max}$ reaches a value of the order of $\approx 1.5$, while in the absence of MHD $\mathrm{Re}\Pi^{(0)}_{max}\approx 0.1$.}
    \label{fig8}
\end{figure}

We investigated the distribution of the screening function $\mathrm{Re}\,\Pi^{(0)}$ on the $q$-$\omega$ plane for two cases: (i) the dispersion has a "Mexican hat" shape, which occurs for $a<\sqrt{2}$, and (ii) there is no MHD, but $a$ is only slightly greater than the critical value, $a=\sqrt{2}$. In the latter case, the band states can be interpreted as intermediate states between the Dirac and Schrödinger states. The results are presented in Fig.~\ref{fig8}. 

First compare the screening functions $\mathrm{Re}\,\Pi^{(0)}(q,\omega)$ for $a=0.1$ (in the presence of MHD) and $a=1.6$ (in absence of MHD). They are shown in panels (a) and (b). Plasmons can be excited in the region above and to the left of the solid line, where $\mathrm{Re}\,\Pi^{(0)}>0$. The function $\mathrm{Re}\,\Pi^{(0)}(q,\omega)$ reaches its maximum on the dashed line. As can be seen, the maximum value of $\mathrm{Re}\,\Pi^{(0)}(q,\omega)$ in the presence of the MHD is approximately 15 times greater than the maximum in the absence of MHD.\@ This ratio also provides an order-of-magnitude estimate of the ratio of the critical values of the interaction parameter $Q=Q_c$ at which IBPs can be excited. In the presence of MHD, this critical value is more than an order of magnitude smaller than in the absence of MHD.\@ 

\begin{figure}
    \centerline{\includegraphics[width=1.0\linewidth]{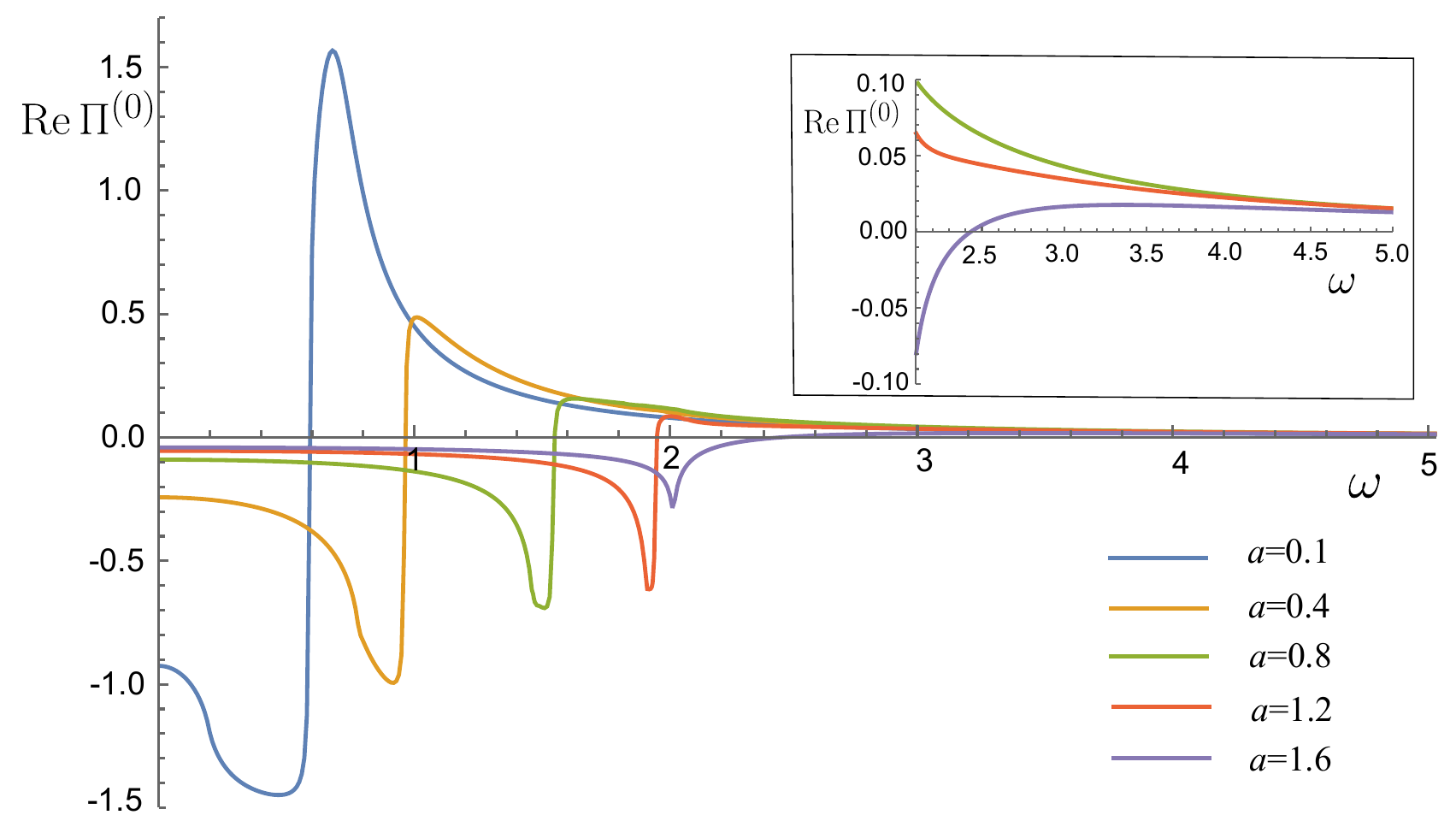}}
    \caption{Screening function $\mathrm{Re}\Pi^{(0)}(\omega)$ at a given wave vector $q=0.3$ for a set of the hybridization parameter $a=0.1 - 1.6$. The inset shows the graphs in the range $2.2<\omega<5$ on an enlarged scale.}
    \label{fig9}
\end{figure}

This conclusion is confirmed by the calculation of $\mathrm{Re}\Pi^{(0)}$ as a function of $\omega$ for a given wave vector, carried out for a wide range of values of the parameter $a$, from 0.1 to 1.6. The results of such calculations are shown in Fig.~\ref{fig9}. As can be seen, the maximum value of $\mathrm{Re}\Pi^{(0)}_{max}$ decreases with increasing $a$ by more than an order of magnitude, from $\mathrm{Re}\Pi^{(0)}_{max}\approx 1.57$ for $a=0.1$ to $\mathrm{Re}\Pi^{(0)}_{max}\approx 0.0177$ for $a=1.6$. Accordingly, one can expect that the critical value of the interaction parameter will increase by almost two orders of magnitude. Of course, this is only a rough estimate showing the general trend, but it is sufficient to conclude that MHD significantly improves the conditions for excitation of IBPs and their stability.

\section{Effect of e-h asymmetry}\label{S_eh_asymmetry}
Now let us examine how the violation of the e-h symmetry affects the properties of IBPs. In this section we will show that the main effect is the appearance of another plasmonic mode, which, however, is excited by a sufficiently strong symmetry breaking of e-h symmetry. This additional mode exists over a wider range of $q$ values, has a significantly lower frequency, and is more strongly attenuated.

The e-h asymmetry is determined by the model parameter $d$, which can take values from -1 to +1. Depending on the sign of $d$, the MHD is more pronounced in the conduction or valence bands. To be specific, we consider the case $d<0$, since it is typically realized in the materials currently being studied. The results for $d>0$ are not qualitatively different.

\begin{figure}
    \centerline{\includegraphics[width=0.9\linewidth]{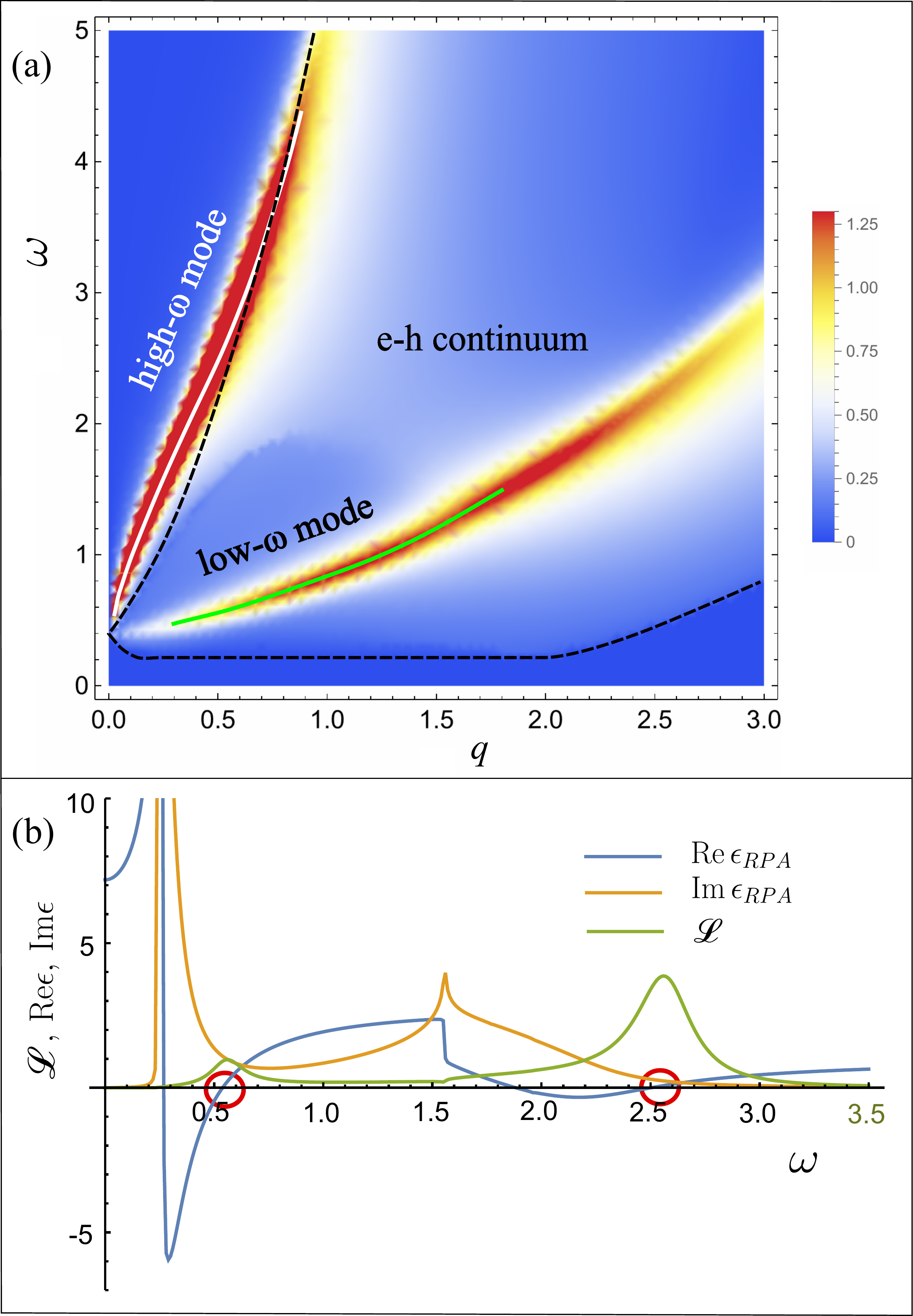}}
    \caption{Energy loss function, IBP spectrum and decay in the case of the MHD with broken e-h symmetry. (a) Distribution of  $\mathcal{L}(q, \omega)$ in the $q$-$\omega$ plane. The black dotted lines indicate the boundaries of the e-h continuum. White line is the dispersion of the high-$\omega$ plasmon mode, and the green line is the dispersion of the low-$\omega$ mode. (b) The real and imaginary parts of the dielectric function $\epsilon_{RPA}$ as functions of frequency for the wave vector $q=0.5$. Red circles show the nodes of $\epsilon_{RPA}(q, \omega)$ for the high-$\omega$ and low-$\omega$ plasmon modes. The corresponding plasmon resonance peaks of the loss function are shown by green line. 
    The calculations were performed for parameters $a=0.2$, $d=-0.8$, $Q=10$.}
    \label{fig10}
\end{figure}

Direct calculations of the dielectric function with using Eqs.~(\ref{eq.dielectric_func}), (\ref{eq.Lidhard_func}) lead to the results illustrated in Fig.~\ref{fig10}. Both modes, high-frequency and low-frequency ones, are clearly seen as corresponding picks of the energy loss function $\mathcal{L}(q, \omega)$. The high-$\omega$ branch originates from a single plasmonic mode in the symmetric case as a result of smooth evolution with increasing $|d|$. The low-$\omega$ branch emerges due to the e-h asymmetry as an overdamped mode at small $|d|$, but the attenuation decreases quite rapidly as $|d|$ becomes sufficiently large. 

The evolution of the main, high-$\omega$, mode with increasing $|d|$ occurs primarily as a smooth shift of the upper boundary of the interval $q$ in which this mode is excited, and an increase of attenuation. The dispersion line of the high-$\omega$ mode $\omega_{p1}(q)$ passes closer to the e-h continuum than in the e-h symmetric case, for comparison see Fig.~\ref{fig5} (a). As a result, the Landau damping of the high-$\omega$ mode is greater than in the symmetric case.

The additional, low-$\omega$, branch of the plasmon spectrum appears at small $|d|$, initially as a strongly damped mode. As $|d|$ increases, the damping decreases, and the excitation becomes quite long-lived. The value of $d$ at which antiscreening occurs, that is, $\mathrm{Re}\,\Pi^{(0)}{(q, \omega)}>0$, depends on the interaction parameter. The dispersion line of the low-$\omega$ mode, $\omega_{p2}(q)$, lies within the region of the e-h continuum and therefore this mode decays much more strongly. An unexpected feature of the additional branch is its low frequency, which can be smaller than the width of the direct interband energy gap.

\begin{figure}
    \centerline{\includegraphics[width=0.9\linewidth]{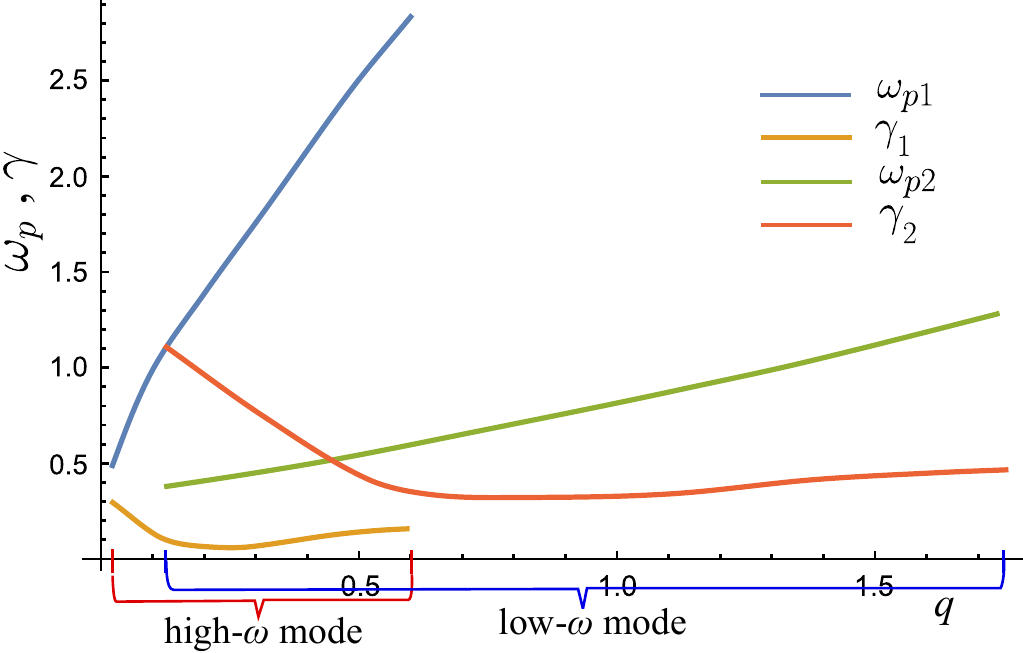}}
    \caption{Dispersion $\omega_{p1,2}(q)$ and the decay constant $\gamma_{1,2}(q)$ of the high-$\omega$ and low-$\omega$ plasmon modes. Red and blue curly brackets on the $q$ axis indicate the ranges in which the high-$\omega$ and low-$\omega$ branches of the plasmon spectrum exist. The calculations were performed for parameters $a=0.2$, $d=-0.8$, $Q=10$.}
    \label{fig11}
\end{figure}

The real and imaginary parts of $\epsilon_{RPA}$ as functions of frequency are presented in Fig.~\ref{fig10} (b) for a fixed $q$, chosen such that both plasmonic modes are present. It is clearly seen that two frequency regions exist where $\mathrm{Re}\,\epsilon_{RPA}<0$ and, therefore, anti-screening can occur. Correspondingly there are two plasmon resonances of the loss function. However, the value of $\mathrm{Im}\,\epsilon_{RPA}$ for the low-$\omega$ mode is significantly larger than that for the high-$\omega$ mode, indicating a significantly stronger damping of the low-$\omega$ mode. The calculated spectrum of both plasmon modes and their decay constants are presented in Fig.~\ref{fig11}.\,\footnote{Regarding Fig.~\ref{fig11}, it should be noted that calculations of $\omega_{p2}$ and $\gamma_2$ are only justified for sufficiently weak attenuation. Therefore, the calculations of the low-$\omega$ mode at $q<0.6$ are only qualitative in nature, but in this region, quantitative results are not very important, since we are interested in plasmons with sufficiently weak attenuation.} 
For the chosen set of parameters, the quality factor of the high-$\omega$ mode reaches 25, and the quality factor of low-$\omega$ mode slightly exceeds 2.

\begin{figure}
    \centerline{\includegraphics[width=1.\linewidth]{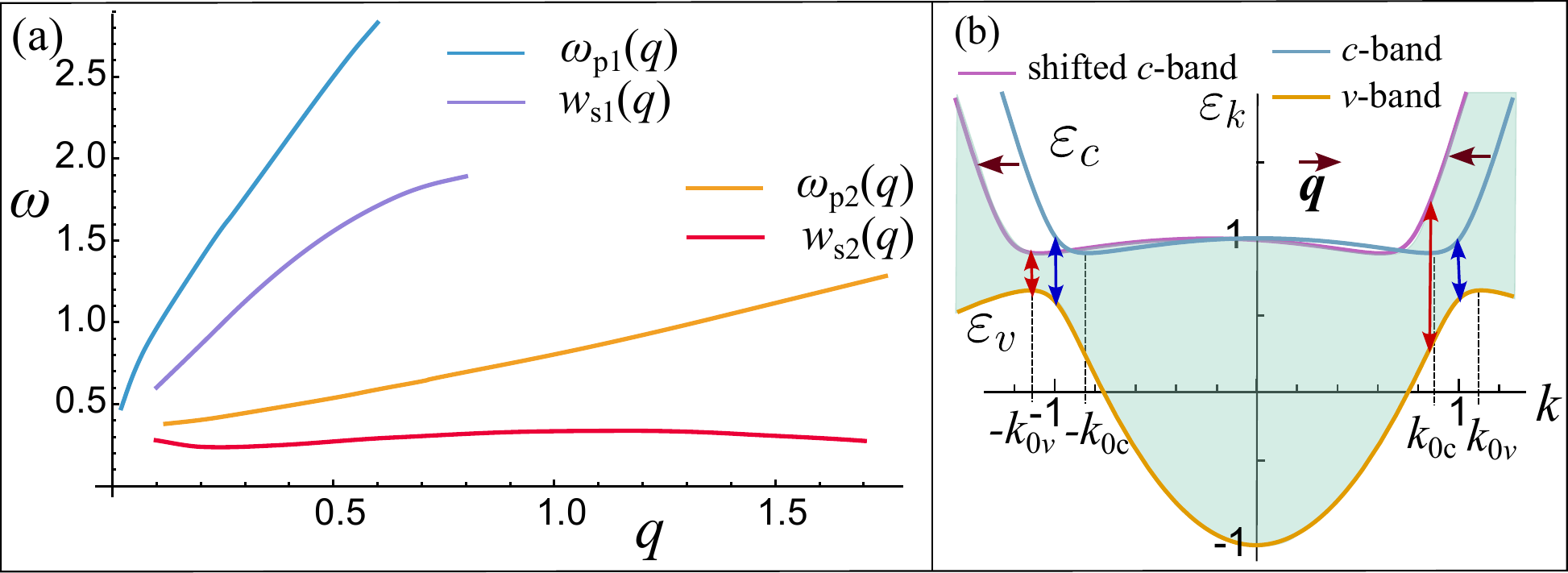}}
    \caption{(a) Spectra of the high-$\omega$ and low-$\omega$ plasmon modes, $\omega_{p1}(q)$ and $\omega_{p2}(q)$, and the energies of the saddle-point singularities, $w_{s1}(q)$ and $w_{s2}(q)$. The calculations were performed for parameters $a=0.2$, $d=-0.8$, $Q=10$. (b) Band diagram explaining the appearance of electron excitations with energy less than the direct interband gap and the anisotropy of the interband energy. Blue arrows show the minimum direct gap between the $c$- and  $v$-bands for excitations with $q\to 0$. For excitations with a finite $q$, the electron transitions can be considered as vertical transitions from the $v$-band into the $c$-band shifted by the vector $\bm{q}$. In this case, the minimum interband gap is shown by red arrows for low- and high-energy excitations. The shaded region represents the interband energy in $k$-space for indirect transitions. It is clear that, the indirect interband gap is anisotropic, and the minimum gap is smaller than the direct gap.}
    \label{fig12}
\end{figure}

The origin of the two collective modes with high and low $\omega$ is easily understood by referring to the singularities of the JDOSq, studied in Sec.\ref{s_JDOS}. When the e-h symmetry is broken, the interband energy becomes a highly anisotropic function in $k$-space as Fig.~\ref{fig3}\,(d-f) shows. In general, there are five singularity points of the JDOSq discussed in Sec.~\ref{s_JDOS}, but only two of them are the most important. These are the high-energy saddle point and low-energy saddle point, the energies of which are presented in Fig.~\ref{fig4}. The high-energy saddle point gives rise to the high-$\omega$ mode, and the low-energy saddle point leads to the low-$\omega$ mode. The correspondence between the plasmon mode spectra and energies of the JDOSq singularities is demonstrated in Fig.~\ref{fig12}\,(a). Quantitatively, the mode frequencies, $\omega_{p1}$ and $\omega_{p2}$, calculated from the zeros of $\epsilon_{RPA}(q, \omega)$, are significantly higher than the energy of the singularities, $w_{s1}$ and $w_{s2}$. The difference is obviously due to the e-e interaction, which increases the frequency of the collective modes, but is absent in the calculation of the interband energy.

The appearance of electron excitations with an energy less than the direct interband gap, due to which the low-$\omega$ branch of the plasmon spectrum is formed, can be graphically explained by the presence of indirect electron transitions in the band diagram shown in Fig.~\ref{fig12}\,(b). These low-energy excitations, and correspondingly the low-$\omega$ branch of plasmons, are present only in the case where the e-h symmetry is absent.

\section{Discussion and concluding remarks}\label{S_conclusion}

We have provided a rather comprehensive theory of the IBP modes in 2D topological insulators with a MHD arising due to the hybridization of inverted electron and hole bands. The theory reveals specific features of the plasmons caused by a bunch of nontrivial properties of this system, such as the two-valued topology of isoenergetic contours, singularities of the density of states, and nontrivial quantum geometric properties of the band states. We have found that in this system there are two physical quantities that play a key role in studying IBPs and understanding their specific properties. These are the JDOSq of the interacting bands for indirect excitations with the wave vector $\bm{q}$ and the overlap function of the spinors describing the multi-orbital structure of the band states, which also depends on $\bm{q}$ and reflects the quantum metric. 

Unlike the usual JDOS with zero wave vector, which in the case of MHD is characterized by the presence of a van Hove singularity on the circle, the JDOSq with finite wave vector has singularities, saddle points and extrema, which are significantly restructured when $\bm{q}$ changes. The interband overlap function has characteristic dips in $k$-space and continuously tends to zero in the limit $q\to 0$. The interaction of the singularities of the JDOSq and the features of the overlap function in $k$-space is an important factor determining the features of the plasmon spectrum specific to the model under study. 

The most important result is that the presence of the MHD strongly facilitates the conditions for excitation of IBPs and strengthens the plasmons. Due to the MHD the plasmonic resonances are significantly enhanced, all other things being equal. Quantitatively, the effect is determined by the critical value of the electron interaction parameter $Q_c$, above which, at $Q>Q_c$, the plasmon modes are excited. $Q_c$ is an important parameter characterizing the role of e-e interaction in the plasmon properties.

In the presence of the MHD, the critical value $Q_c$ is orders of magnitude lower than in the absence of MHD. Physically, this effect is caused by a strong increase in the polarizability of the electron system due to electron transitions between two different isoenergetic contours present in both the valence and conduction bands. The effect is stronger, the deeper the MHD profile. A decrease in $Q_c$ leads to a significant increase in the intensity of the plasmon resonance with the same strength of the e-e interaction and a decrease in damping.

In addition, a number of unusual features of the plasmon spectrum related to the JDOSq and the quantum metric were discovered. 

Plasmon excitations arise in a limited range of the wave vector $q_l<q<q_u$. The lower boundary $q_l$ arises from the combined action of two factors: the tendency of the interband overlap function to zero as $q\to 0$, and the reconstruction of van Hove singularities of the JDOSq as $q\to 0$. The upper boundary $q_u$ is due to strong Landau damping at large $q$.

The spectrum of plasmon modes is closely related to the spectrum of the singularities of the JDOSq which depend on $q$. Among all singularities that appear in the general case, only saddle points play a key role since they give rise to plasmon modes. However, the plasmon frequency $\omega_p$ exceeds the singularity energy $w_s$ by an amount determined by the e-e interaction.

When an electron system has e-h symmetry, there is a single plasmonic mode. Increasing the interaction strength leads to an increase in the frequency $\omega_p(q)$ and a decrease in the damping $\gamma(q)$. Quantitatively, increasing the interaction parameter to $Q=10$ with its critical value $Q_c\approx 1$ results in a nearly twofold increase in frequency and an increase in the quality factor to value of the order of $10^2$.

It is important to note that in the absence of MHD in the BHZ model with the same parameters but a slightly larger hybridization parameter $a>\sqrt{2}$, plasmons can also be excited, but this requires a significantly stronger interaction. Thus, as the hybridization parameter increases from $a=1.2$ (MHD is present) to $a=1.6$ (MHD is absent), the critical value of the interaction parameter increases by almost an order of magnitude.

The most nontrivial effect arises from the breaking of the e-h symmetry, which leads to the appearance of another branch of the interband excitation spectrum in addition to the main mode. This additional mode has a significantly lower frequency and stronger damping. The emergence of this mode is explained by a radical rearrangement of the singularities of the JDOSq, which occurs due to the breaking of e-h symmetry. As a result, two saddle-point singularities appear, the energy gap between which increases with increasing $q$. Each of the singularities generates a branch of the plasmon spectrum. For a small parameter of e-h asymmetry $|d|\ll 1$, the additional mode is overdamped, but as $|d|$ increases, the damping decreases quite rapidly.

The numerical evaluations presented in this paper were performed for typical parameters of topological insulators such as HgTe/CdHgTe quantum wells~\cite{BHZ,krishtopenko2016phase}, inverted InAs/GaSb quantum wells~\cite{PhysRevLett.107.136603} strained-layer InAs/In$_x$Ga$_{1-x}$Sb quantum wells~\cite{PhysRevResearch.4.L042042,Zhang_2025}. In this case the MHD can be realized in the topological phase. To be specific, for the numerical calculations presented in this article, the material parameters $a$, $d$, and $Q$, key in the developed theory, were chosen to be close to the parameters of InAs/In$_x$Ga$_{1-x}$Sb quantum wells given in Refs.~\cite{PhysRevResearch.4.L042042,Zhang_2025}.

There are also many materials with MHD of a different nature, the electron system of which has largely similar properties. In this regard, let us focus on the properties that are necessary for the emergence of the main features of IBPs described here.

The formation of a two-valued structure of isoenergetic contours in the $c$- and $v$-bands and the associated enhancement of electronic polarizability appears to be a common property of many materials with MHD. The critical parameters of the system for observing the described effects are the depth of the Mexican hat shape and the interaction parameter $Q$. Within the BHZ model, the shape depth is controlled by the parameter $a$. Clearly, the effects studied here are most pronounced for deep shape of the Mexican hat, where intercontour transitions of electrons make a significant contribution to polarizability. Moreover, it does not matter in which of the bands the dispersion has a Mexican hat shape. The magnitude of the interaction parameter is determined by the ratio of the Coulomb energy of an electron at a characteristic distance to the interband energy. In the BHZ model, the characteristic distance is $\sqrt{|B/M|}$ and the energy scale is $|M|$. Since the physical meaning of the critical parameters is quite universal, it can be assumed that the effects studied here may also manifest themselves in other MHD models.

In conclusion, we reached the following key findings:\\
\indent i) The presence of a MHD significantly strengthens interband plasmons, facilitates the conditions for plasmon generation, and increases their frequency and quality factor.\\
\indent ii) The strengthening of plasmons is due to a strong increase in polarizability of the electron fluid because of electron transitions between different isolines of interband energy.\\
\indent iii) The plasmon spectrum is confined within a finite range of the wave vector with lower and upper boundaries $q_l$ and $q_u$, which depend on the strength of the e-e interaction.\\
\indent iv) The role of e-e interaction in the formation of the width of the excitation region of plasmons in $q$-space, their frequency and quality factor is characterized by the excess of the interaction parameter $Q$ over its critical value $Q_c$, which is significantly reduced due to MHD.\\
\indent v) Breaking the e-h symmetry leads to the appearance of an additional branch of the plasmon spectrum with a lower frequency and stronger damping.\\
\indent vi) The unusual properties of plasmons can be explained by the combined action of two factors: a strong reconstruction of the van Hove singularities of the JDOSq when the plasmon wave vector changes and the features of the interband overlap function, which also changes depending on $q$.

\begin{acknowledgments}
This work was carried out in the framework of the state task (theme code FFWZ-2025-0014) for the Kotelnikov Institute of Radio Engineering and Electronics.
\end{acknowledgments}
        
\bibliography{ibp_paper}

\end{document}